\documentclass[aps,twocolumn,amsmath,amssymb]{revtex4-2}
\usepackage[dvipdfmx]{graphicx} 
\usepackage{bm}
\usepackage{braket}
\usepackage{dcolumn}
\usepackage{hyperref}
\hypersetup{colorlinks=true, urlcolor=blue, linkcolor=blue, citecolor=blue}

\begin{document}

\title{Electronic band structure, phonon dispersion, and magnetic triple-\textit{q} state in GdGaI}

\author{Tatsuya Kaneko,$^1$ Ryota Mizuno,$^2$ Shu Kamiyama,$^1$ Hideo Miyamoto,$^1$ and Masayuki Ochi$^{1,2}$}

\affiliation{$^1$Department of Physics, The University of Osaka, Toyonaka, Osaka 560-0043, Japan\\
$^2$Forefront Research Center, The University of Osaka, Toyonaka, Osaka 560-0043, Japan}

\date{\today}

\begin{abstract}
We theoretically investigate the physical properties of the magnetic van der Waals material GdGaI. Using first-principles calculations, we compute the phonon dispersion of GdGaI and show no imaginary phonons, suggesting that phonon-driven phase transitions are unlikely to occur in GdGaI. Our band calculation reveals that the electronic bands near the Fermi energy are composed of Gd $5d$ and Ga $4p$ orbitals. We construct a tight-binding model that incorporates the Gd $5d$ and Ga $4p$ orbitals to investigate the magnetic structure. We introduce Kondo coupling between electrons in Gd $5d$ orbitals and localized spins in Gd $4f$ orbitals and present the modified band structure when localized spins form a magnetic order characterized by three $\bm{q}$ vectors that connect the valence and conduction bands. We discuss the origin of the spin order based on the Ruderman-Kittel-Kasuya-Yosida mechanism and suggest that Coulomb interactions acting on electrons near the Fermi level can contribute to the ordering of localized spins. 
\end{abstract}

\maketitle

\section{Introduction}

Van der Waals (vdW) materials have played a key role in the development of modern materials science~\cite{AGeim2007,AGeim2013,MChhowalla2013}. 
For instance, atomically thin vdW materials have provided platforms for two-dimensional (2D) physics~\cite{ACastroNeto2009,KMak2016,SManzeli2017,GWang2018}, and stacking engineering has enabled the creation of reorganized material properties~\cite{YCao2018,LBalents2020,EAndrei2020,LWang2020}. 
Layered materials with spin orders, so-called vdW magnets, have recently attracted attention~\cite{KBurch2018,SYang2021,HKurebayashi2022,QWang2022}. 
For example, layer-number-dependent spin structures and excitons coupled with background magnetism have been reported in various vdW magnets~\cite{BHuang2017,SKang2020,JSon2021,SSon2022,MZiebel2024}. 
Studying vdW magnets will be important for developing spintronic materials and devices. 

\begin{figure}[t]
\begin{center}
\includegraphics[width=0.88\columnwidth]{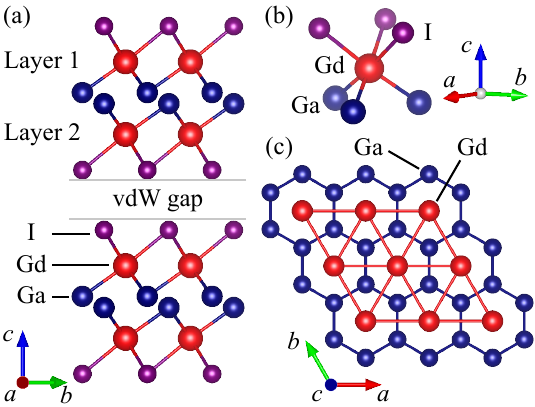} 
\caption{(a) Side view of the crystal structure of GdGaI visualized by VESTA~\cite{VESTA}. 
Red, navy, and purple balls represent Gd, Ga, and I atoms, respectively.
(b) GdGa$_3$I$_3$ octahedron. 
(c) Top view of the Gd and Ga sites.} 
\label{fig1}
\end{center}
\end{figure}

GdGaI is a recently reported magnetic vdW material~\cite{ROkuma_arXiv}. 
GdGaI is isostructural to CeSiI~\cite{ROkuma2021,VPosey2024}. 
Figure~\ref{fig1}(a) shows the side view of the crystal structure. 
Three Ga and three I atoms are octahedrally coordinated around a Gd atom [Fig.~\ref{fig1}(b)], and their bilayer units are stacked in the $c$-axis direction. 
The outer layers that face the vdW gaps are structured by I. 
The Gd atoms make a bilayer triangular lattice, while the inner Ga atoms make a buckled honeycomb lattice, as shown in Fig.~\ref{fig1}(c). 
Angle-resolved photoemission spectroscopy (ARPES) unveils the characteristic electronic band structure of GdGaI~\cite{ROkuma_arXiv}. 
The band structure at high temperatures appears to be semiconducting with a narrow gap, where the valence band (VB) top and conduction band (CB) bottoms are at the $\Gamma$ and M points, respectively, in the Brillouin zone (BZ). 
The gap size increases as temperature decreases, and the VB top becomes flat (or camelback shape) around the $\Gamma$ point at low temperatures. 
In addition, spectral weights reflecting the replica of the flattened VB top emerge around the M point. 
This temperature-dependent ARPES spectrum is similar to the spectral property of TiSe$_2$~\cite{TKidd2002,HCercellier2007,TRohwer2011,KSugawara2016}, and the possibility of a transition to an excitonic insulator state~\cite{DJerome1967,BHalperin1968,TKaneko2025} has been discussed in GdGaI~\cite{ROkuma_arXiv}. 
In contrast to TiSe$_2$, Gd $4f$ orbitals host localized $S=7/2$ spins, and GdGaI shows a magnetic phase transition near 30~K. 
Based on data from nuclear magnetic resonance, Ref.~\cite{ROkuma_arXiv} suggests the triple-$\bm{q}$ magnetic order with scalar spin chirality in the low-temperature phase, where the order is characterized by three $\bm{q}$ vectors that connect the $\Gamma$ and M points. 
While GdGaI has exhibited fascinating material properties in experiments, its electronic, structural, and magnetic properties have not been comprehensively studied by theoretical calculations. 

In this paper, we investigate the fundamental physical properties of GdGaI using first-principles calculations and a tight-binding (TB) model with Kondo coupling. 
First, we present the phonon dispersion of GdGaI and show no phonon softening, suggesting that a structural phase transition is unlikely to occur through a phonon-driven mechanism alone. 
Our band calculation reveals that the electronic band structure near the Fermi level is composed of Gd $5d$ and Ga $4p$ orbitals. 
Then, we construct a TB model that incorporates the Gd $5d$ and Ga $4p$ orbitals to investigate the magnetic structure in GdGaI. 
We introduce Kondo coupling between electrons in Gd $5d$ orbitals and localized spins in Gd $4f$ orbitals (i.e., we consider the Kondo lattice model) and evaluate the band structure when localized spins form a triple-$\bm{q}$ order with scalar spin chirality. 
Finally, we discuss the origin of the spin order based on the Ruderman-Kittel-Kasuya-Yosida (RKKY) mechanism~\cite{MRuderman1954,TKasuya1956,KYoshida1957,SHayami2021} and suggest that Coulomb interactions acting on $d$ electrons near the Fermi level can contribute to the ordering of localized spins on $f$ orbitals. 

The rest of this paper is organized as follows.
In Sec.~\ref{sec_II}, we present the results of our first-principles calculations, including the phonon and electronic band dispersions. 
In Sec.~\ref{sec_III}, we construct a TB model composed of the Gd $5d$ and Ga $4p$ orbitals. 
Then, we introduce Kondo coupling in the TB model and present the band structure in the triple-$\bm{q}$ order. 
We also discuss the origin of the magnetic order based on the RKKY mechanism. 
A summary and perspective are presented in Sec.~\ref{sec_IV}.

\section{First-principles calculations} \label{sec_II}

\subsection{Methods}

First-principles calculations based on density functional theory (DFT) were performed using the projector-augmented wave (PAW) method~\cite{PEBlochl1994} as implemented in the Vienna {\it ab initio} simulation package (VASP)~\cite{GKresse1993,GKresse1994,GKresse1996,GKresse1996_2}. 
We used the Heyd-Scuseria-Ernzerhof (HSE) hybrid functional~\cite{JHeyd2003,JHeyd2004,AVKrukau2006} with the range-separation parameter $\mu=0.15$~\AA$^{-1}$ to obtain the appropriate band structure as discussed later. 
We also used the generalized gradient approximation with the Perdew-Burke-Ernzerhof (PBE) parametrization~\cite{JPPerdew1996} for comparison.
The results using PBE are presented in Appendix~\ref{appendix_A}. 
Open-core treatment was applied to Gd-$4f$ localized states. 
For Gd, Ga, and I atoms, [Xe $4f^7$]-core, [Ar]-core, and [Kr $4d^{10}$]-core PAW potentials were used, respectively. 
Spin-orbit coupling was not included for simplicity unless otherwise noted. 
Our DFT calculations were performed in the absence of any magnetic order. 
Plane-wave cutoff energy of 400~eV and an $8\times 8\times 3$ $k$ mesh were used. 
The experimental crystal structure taken from Ref.~\cite{ROkuma_arXiv} was used for our DFT calculations except for phonon calculations. 
For phonon calculations, we first optimized the atomic coordinates while the lattice constants $a=4.1985$~\AA \ and $c=11.476$~\AA \ taken from Ref.~\cite{ROkuma_arXiv} were fixed until the Hellmann--Feynman force for each atom became less than $10^{-4}$~eV~\AA$^{-1}$. 
The optimized atomic coordinates are presented in Appendix~\ref{appendix_A}. 
Frozen phonon calculation implemented in the PHONOPY package~\cite{ATogo2023,ATogo2023_JPSJ} was performed using a $3\times 3\times 2$ $q$ mesh.
All calculations were performed for both PBE and HSE. 

\begin{figure}[t]
\begin{center}
\includegraphics[width=\columnwidth]{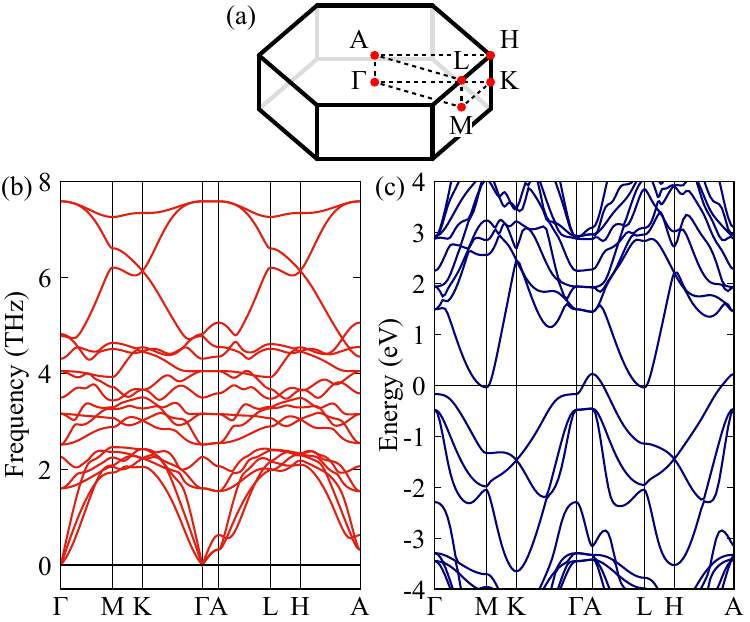} 
\caption{(a) Schematic figure of the BZ. 
(b) Phonon dispersion and (c) electronic band structure obtained with the HSE hybrid functional.
The horizontal line in (c) represents the Fermi energy $E_F$.} 
\label{fig2}
\end{center}
\end{figure}

\subsection{Phonon and nonmagnetic electronic bands} 

\begin{figure*}[t]
\begin{center}
\includegraphics[width=2\columnwidth]{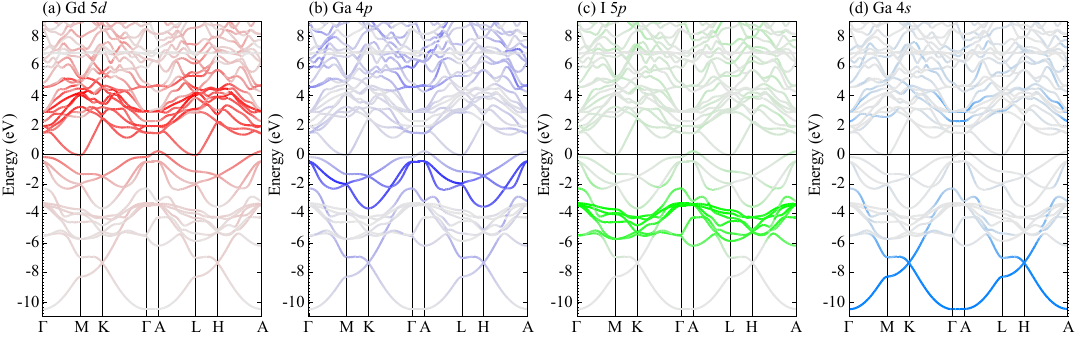} 
\caption{Electronic band dispersions with colored weights: (a) Gd $5d$, (b) Ga $4p$, (c) I $5p$, and (d) Ga $4s$ orbitals, respectively.} 
\label{fig3}
\end{center}
\end{figure*}

Figure~\ref{fig2}(b) presents the phonon dispersion of bulk GdGaI. 
Here, the phonon calculation was performed using the crystal structure optimized with the HSE functional (Table~\ref{table1} in Appendix~\ref{appendix_A}). 
As shown in Fig.~\ref{fig2}(b), the phonon dispersion shows no imaginary phonons, suggesting that the crystal structure of GdGaI is stable. 
Figure~\ref{fig2}(c) plots the electronic band structure. 
Note that the crystal structure experimentally determined in Ref.~\cite{ROkuma_arXiv} was used in the electronic band calculation. 
The VB top and CB bottom are slightly overlapped at different $\bm{k}$ points, and the DFT calculation shows a semimetallic band structure in bulk GdGaI. 
The hole pocket is located around the A point, and the electron pocket resides around the M--L line. 
Figure~\ref{fig2} presents the phonon and electronic band dispersions using the HSE hybrid functional. 
As shown in Appendix~\ref{appendix_A}, even if we employ the PBE functional and the semimetallic band overlap is larger than that in Fig.~\ref{fig2}(c), the phonon calculation does not exhibit an imaginary phonon. 
Therefore, the use of the HSE hybrid functional is not the cause of no phonon softening. 
The VB around the Fermi level is dispersed along the $\Gamma$--A line, and the band structure becomes semimetallic. 
However, note that Ref.~\cite{ROkuma_arXiv} suggests semiconducting electronic properties in the nonmagnetic phase. 

Figure~\ref{fig3} shows the band structures with the weights of the Gd $5d$, Ga $4p$, I $5p$, and Ga $4s$ orbitals. 
The electronic bands near the Fermi energy $E_F$ are mainly composed of the Gd $5d$ and Ga $4p$ orbitals. In contrast, the weights of the I $5p$ and Ga $4s$ orbitals are distributed in the low-energy regions away from $E_F$. 
Hence, the contributions from the I $5p$ and Ga $4s$ orbitals are less significant near $E_F$, and we mainly discuss the contributions of the Gd $5d$ and Ga $4p$ orbitals. 
As seen in Fig.~\ref{fig3}(b), the Ga $4p$ orbitals contribute to the formation of three VBs near $E_F$. 
However, the highest VB contains a substantial weight of the Gd $5d$ orbitals around the $\Gamma$ and A points, suggesting strong $d$-$p$ hybridization between the Gd and Ga sites. 
Due to the honeycomb network of Ga sites, the electronic bands around the K point composed of the Ga $4p$ and $4s$ orbitals form Dirac-like dispersions around $-1.5$~eV and $-7.5$~eV below $E_F$. 

The nonmagnetic band structure of GdGaI is similar to the semimetallic band structure of 1$T$-TiSe$_2$~\cite{TKidd2002,HCercellier2007,TRohwer2011,KSugawara2016}, in which the VB top and CB bottom are at the $\Gamma$ and M (L) points, respectively. 
However, there are several differences between GdGaI and TiSe$_2$. 
First, TiSe$_2$ has six VBs based on the Se $4p$ orbitals below $E_F$~\cite{AZunger1978,YYoshida1980,CFang1997}, indicating that the $p$ orbitals are almost fully occupied. 
However, the number of Ga $4p$-based bands below $E_F$ in GdGaI is three [see Fig.~\ref{fig3}(b)], where nearly half of the Ga $4p$ orbitals are unoccupied. 
This suggests that the energy levels of the Gd $5d$ and Ga $4p$ orbitals are comparable, in contrast to the hierarchy of TiSe$_2$, where the energy level of the Ti 3$d$ orbitals is higher than that of the Se 4$p$ orbitals, and they form the CBs and VBs, respectively. 
Second, the component of the Gd $5d$ orbitals is sizable in the VB top of GdGaI [see Fig.~\ref{fig3}(a)], whereas the weight of the Ti $3d$ orbitals in the VBs of TiSe$_2$ is less significant~\cite{RBianco2015,TKaneko2018,CChen2018}. 
This suggests that the channels of the driving forces for the phase transitions can be different. 
Because the components of the VBs differ in TiSe$_2$ and GdGaI, coupling between the VBs around the $\Gamma$ (A) point and the CBs around the M (L) point for the phase transition should be the $d$-$p$ channel in TiSe$_2$, while that can be the $d$-$d$ channel in GdGaI. 
In TiSe$_2$, the phase transition accompanied by lattice distortion~\cite{FDiSalvo1976} has been explained by $d$-$p$ coupling that involves electron-phonon interactions~\cite{RBianco2015,TKaneko2018,CChen2018}. 
On the other hand, in GdGaI, $d$-$d$ coupling, arising from Kondo (electron-spin) coupling and $d$-$d$ Coulomb interaction, can contribute to the magnetic phase transition. 
We will discuss their roles later. 
Third, the first-principles calculation of TiSe$_2$ shows phonon softening~\cite{MCalandra2011,MHellgren2017} corresponding to the structural phase transition experimentally detected~\cite{FDiSalvo1976}, whereas the phonon dispersion of GdGaI in Fig.~\ref{fig2}(b) exhibits no imaginary phonons. 
This implies that a structural phase transition is unlikely to occur in GdGaI. 
This difference is likely due to the different compositions of the electronic bands near $E_F$, as pointed out above.

\section{Tight-binding model coupled with localized spins} \label{sec_III}

In this section, we examine the roles of localized spins on Gd $4f$ orbitals using the Kondo lattice model. 
For the electronic part of the Kondo lattice model, we construct a TB model based on the Gd $5d$ and Ga $4p$ orbitals. 
The first-principles calculation for bulk GdGaI exhibits the band dispersion along the $\Gamma$--A line ($k_z$ direction) at $E_F$ [see Fig.~\ref{fig2}(c)], whereas semiconducting electronic properties and less significant $k_z$ dispersion are suggested in Ref.~\cite{ROkuma_arXiv}. 
According to our first-principles calculation, the Gd $5d$-based orbitals elongated along the $c$ axis contribute to the $k_z$ dispersion near $E_F$, see Appendix~\ref{appendix_B} for details. 
The consistency with experimental results regarding the presence of the $k_z$ dispersion is currently beyond the scope of our theoretical work. 
Here, we focus on investigating intralayer magnetic order and consider a TB model in a single bilayer [(I--Gd--Ga)--(Ga--Gd--I) layer]. 
First, we extract the hopping parameters for our TB model from the DFT bands. 
Then, we introduce Kondo coupling into the TB model and discuss the electronic bands when localized spins form the magnetic order suggested in Ref.~\cite{ROkuma_arXiv}. 
Finally, we examine the origin of the magnetic order based on the RKKY mechanism. 
Here, we also consider the effects of local Coulomb interactions acting on $d$ electrons.

\subsection{Tight-binding model}

For 2D-layer calculations, we added $\sim$15~\AA\ vacuum layer for the experimental crystal structure of bulk in Ref.~\cite{ROkuma_arXiv}. 
The electronic bands using the HSE hybrid functional with the range-separation parameter $\mu=0.15$~\AA$^{-1}$ are employed as the reference DFT bands, where we used a $12\times 12 \times 1$ $k$ mesh. 
The band structures calculated using PBE and HSE with different values of the range-separation parameter $\mu$ are presented in Appendix~\ref{appendix_C}. 
The band structure obtained by the PBE functional is semimetallic.
The band gap opens as $\mu$ decreases (see Appendix~\ref{appendix_C}), and we employ the nearly zero-gap band structure obtained by the HSE hybrid functional with $\mu=0.15$~\AA$^{-1}$. 
As seen in Fig.~\ref{fig3}, the Gd $5d$ and Ga $4p$ orbitals compose the VBs and CBs near $E_F$. 
To construct a TB model, we extracted maximally localized Wannier orbitals~\cite{NMarzari1997,ISouza2001} of the Gd-$5d$ and Ga-$4p$ states using the WANNIER90 code~\cite{GPizzi2020}. 
The outer and inner energy windows for Wannierization were [$-3.0$:8.0]~eV and [$-1.6$:2.7]~eV, respectively. 
We employ five Gd $5d$ orbitals and three Ga $4p$ orbitals in the top and bottom layers, where we consider 16 (=~$5\times2+3\times 2$) orbitals in total for Wannierization.

\begin{figure}[b]
\begin{center}
\includegraphics[width=\columnwidth]{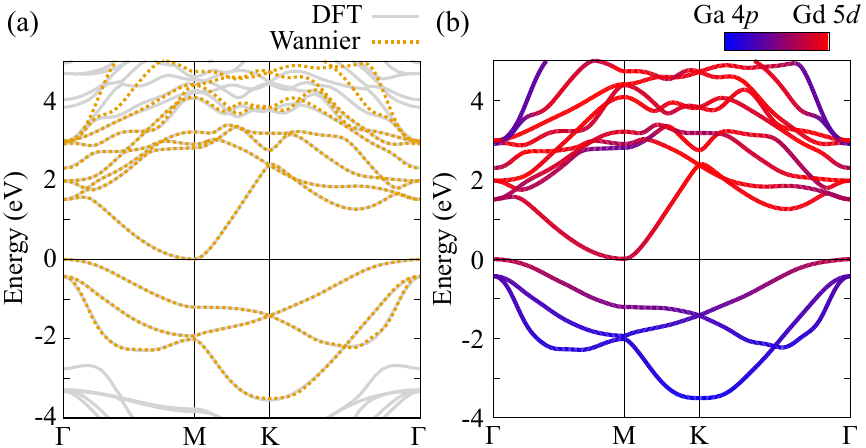} 
\caption{(a) Band structure of the 16-orbital $d$-$p$ model obtained by Wannierization (orange dotted lines), where the modeled band structure is superposed on the first-principles band structure using the HSE hybrid functional (gray lines). 
(b) Electronic bands colored by the orbital components in the $d$-$p$ model. 
Red and blue indicate the Gd $5d$ and Ga $4p$ orbitals, respectively.} 
\label{fig4}
\end{center}
\end{figure}

Figure~\ref{fig4}(a) compares the DFT bands with the TB bands obtained by Wannierization. 
In single-bilayer GdGaI, the VB top and CB bottom are located at the $\Gamma$ and M points, respectively. 
As shown in Fig.~\ref{fig4}(a), the electronic bands of the TB model are in good agreement with the DFT bands near $E_F$. 
Figure~\ref{fig4}(b) shows the TB bands with the weight of the Gd $5d$ component. 
The orbital components of the DFT calculation [e.g., Fig.~\ref{fig3}(a)] are well reproduced in our TB model. 

\subsection{Coupling with localized spins} \label{sec_IIIB}

We consider the Kondo coupling term on the TB model obtained in the previous subsection to discuss the relationship between the magnetic order of localized spins and the electronic band structure near $E_F$. 
In this subsection, we consider the Hamiltonian that comprises $\hat{H} = \hat{H}^{(0)}_{dp} + \hat{H}_{\rm K}$. 
The electronic bands in the TB model are described by 
\begin{align}
\hat{H}^{(0)}_{dp} = \sum_{j,j'} \sum_{\xi,\xi'} t_{j\xi,j'\xi'} \hat{c}^{\dag}_{j,\xi} \hat{c}_{j',\xi'}, 
\label{eq:Ham_0_dp}
\end{align}
where $\xi = (\Lambda,l,\sigma)$ includes the indices of atom $\Lambda$ [= Gd(1), Gd(2), Ga(1), Ga(2)], orbital $l$, and spin $\sigma$. 
$j$ represents position $\bm{r}_j$, which is given by the translation vectors $\bm{a}_1 = a (1,0)$ and $\bm{a}_2 = a (-1/2,\sqrt{3}/2)$ (where $a$ is the lattice constant). 
$t_{j\xi,j'\xi'}$ is the transfer integral between $(j,\xi)$ and $(j',\xi')$, which includes the energy level $\epsilon_{\xi}$ of $\xi$. 
For the Gd and Ga atoms, 
\begin{align}
\hat{c}_{j,\xi} = 
\left\{
\begin{array}{ll}
\hat{d}_{j,\gamma,m,\sigma} & \left[\, \Lambda = {\rm Gd}(\gamma) \, \right] \\
\hat{p}_{j,\gamma,n,\sigma} & \left[\, \Lambda = {\rm Ga}(\gamma) \, \right]
\end{array}
\right. , 
\end{align}
where $\hat{d}_{j,\gamma,m,\sigma}$ and $\hat{p}_{j,\gamma,n,\sigma}$ are the annihilation operators of an electron for the Gd $5d$ and Ga $4p$ orbitals, respectively, and $\gamma$ (=1,2) is the layer index. 
The orbital indices in the Gd and Ga sites are represented by $m$ and $n$, respectively. 
$\hat{H}_{\rm K}$ is the Kondo coupling term 
\begin{align}
\hat{H}_{\rm K} = -J_{\rm K} \sum_{j} \sum_{\gamma} \sum_{m} \sum_{\sigma,\sigma'} \left( \hat{d}^{\dag}_{j,\gamma, m,\sigma} \bm{\sigma}_{\sigma\sigma'} \hat{d}_{j,\gamma, m,\sigma'} \right) \cdot \hat{\bm{S}}_{j,\gamma},
\label{eq:H_K}
\end{align}
where $\bm{\sigma} = (\sigma^x, \sigma^y, \sigma^z)$ is the vector of Pauli matrices and $\hat{\bm{S}}_{j,\gamma}$ represents a localized spin that models the electron spin on the Gd $4f$ orbitals at site $j$ and layer $\gamma$. 
$J_{\rm K}$ is the coupling constant between the $d$ electrons and localized spin. 

\begin{figure}[t]
\begin{center}
\includegraphics[width=\columnwidth]{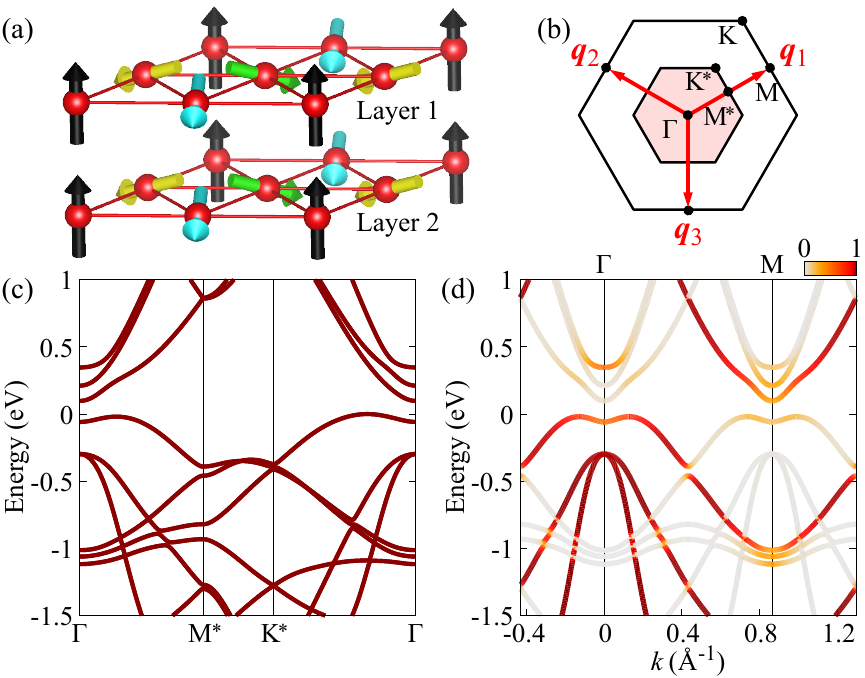} 
\caption{(a) All-out structure of localized spins. 
Red balls represent Gd atoms. 
Arrows represent localized spins on Gd $4f$ orbitals. 
(b) BZ and three $\bm{q}$ vectors that connect the $\Gamma$ and M points. 
(c) Band structure with the all-out spin structure in (a), where the electronic bands for $J_{\rm K}=0.4$~eV are plotted in the reduced BZ [shadowed area in (b)]. 
(d) Band structure along the $\Gamma$--M line in the original BZ, where color indicates the weight of the $\bm{q}=\bm{0}$ component.} 
\label{fig5}
\end{center}
\end{figure}

Here, we treat localized spins as vector spins, i.e., $\hat{\bm{S}}_{j,\gamma} \rightarrow \bm{S}_{j,\gamma}$, and discuss electronic band structures with ordered spins. 
Since the VB top is located at the $\Gamma$ point and the CB bottoms are located at the M points, a spin order may be characterized by the $\bm{q}$ vectors
\begin{align}
\bm{q}_{1} \! = \! \left( \frac{\pi}{a}, \frac{\pi}{\sqrt{3}a} \right), \;
\bm{q}_{2} \! = \! \left(-\frac{\pi}{a}, \frac{\pi}{\sqrt{3}a} \right), \;
\bm{q}_{3} \! = \! \left( 0,- \frac{2\pi}{\sqrt{3}a} \right), 
\label{eq:three_q}
\end{align}
which connect the $\Gamma$ and three M points as shown in Fig.~\ref{fig5}(b). 
Using these $\bm{q}$ vectors, the triple-$\bm{q}$ spin structure is described by 
\begin{align}
\bm{S}_{j,\gamma} = \sum^3_{\alpha=1} \bm{S}_{\alpha,\gamma} \cos(\bm{q}_{\alpha} \cdot \bm{r}_j). 
\label{eq:S_triple_q}
\end{align} 
Here, we consider the all-out spin structure that possesses scalar spin chirality suggested by the experiment in Ref.~\cite{ROkuma_arXiv}. 
The all-out spin structure shown in Fig.~\ref{fig5}(a) is realized by 
\begin{align}
\bm{S}_{\alpha,\gamma} = \frac{S}{\sqrt{3}} \eta_{\gamma} \bm{e}_{\alpha},
\end{align}
with the unit vectors $\bm{e}_1 = (1/\sqrt{2}, 1/\sqrt{6}, 1/\sqrt{3})$, $\bm{e}_2 = (-1/\sqrt{2}, 1/\sqrt{6}, 1/\sqrt{3})$, and $\bm{e}_3 = (0, - 2/\sqrt{6}, 1/\sqrt{3})$, where $|\bm{S}_{j,\gamma}|=S$ in Eq.~(\ref{eq:S_triple_q}). 
Note that net magnetization is zero in the spin structure we set. 
The spin structure in one layer is equivalent to the all-out spin structure studied in the triangular Kondo lattice model~\cite{IMartin2008,YAkagi2010,YKato2010}. 
$\eta_{\gamma}$ is the sign, where the spins in layers 1 and 2 are parallel when $\eta_1=\eta_2=1$ while they are antiparallel when $\eta_1=-\eta_2=1$. 
Figure~\ref{fig5}(a) is the spin structure with $\eta_1=\eta_2=1$ (see also Appendix~\ref{appendix_D}). 
While $\eta_1=\eta_2=-1$ forms the all-in spin structure~\cite{ROkuma_arXiv}, it gives the equivalent band dispersion as the all-out case when net magnetization is zero. 
Since the unit cell is extended to $2\times 2$ in the triple-$\bm{q}$ spin structure, the Hamiltonian is diagonalized in the reduced BZ (shadowed area) shown in Fig.~\ref{fig5}(b). 
When $J_{\rm K} \ne 0$, band hybridization of $\bm{k}$ and $\bm{k}+\bm{q}_{\alpha}$ components can occur in the reduced BZ because of the off-diagonal elements proportional to $J_{\rm K}$. 
In calculations with spin orders, we set $|\bm{S}_{j,\gamma}|=S=1$, where the value of $J_{\rm K}$ used in our band calculation should be considered to include the amplitude of spin. 

Figure~\ref{fig5}(c) presents the electronic band structure under the all-out spin structure with $\eta_1=\eta_2 = 1$ shown in Fig.~\ref{fig5}(a). 
The CB bottoms at three M points are folded to the $\Gamma$ point in the reduced BZ (see also Fig.~\ref{fig10} in Appendix~\ref{appendix_C}). 
Then, the VB and CB are hybridized due to the Kondo coupling $J_{\rm K}$. 
Since both VB top and CB bottom contain the Gd $d$-orbital components, the triple-$\bm{q}$ order of localized spins can lead to the gap opening due to $d$-$d$ hybridization via Kondo coupling. 
Because of this hybridization gap opening, the VB top around the $\Gamma$ point becomes flat, similarly to the ARPES spectra at low temperatures~\cite{ROkuma_arXiv}. 
The electronic band structure with $\eta_1=-\eta_2 = 1$ is presented in Appendix~\ref{appendix_D}, where the antiparallel spin structure does not open a noticeable gap near $E_F$. 
Hence, the energy of the electronic part under the spin state with $\eta_1=\eta_2$ (parallel) is lower than that with $\eta_1=-\eta_2$ (antiparallel). 
The energies of the triple-$\bm{q}$ and single-$\bm{q}$ spin orders are competitive. 
In our TB model with Kondo coupling, the energy of the triple-$\bm{q}$ order shown in Fig.~\ref{fig5}(a) is slightly lower than that of the single-$\bm{q}$ order. 
Among possible short-period spin structures characterized by $\bm{q}_{\alpha}$ in Eq.~(\ref{eq:three_q}), the triple-$\bm{q}$ spin order with $\eta_1=\eta_2$ is energetically stable in our TB model. 

To discuss the spectral weight around the M point observed in ARPES~\cite{ROkuma_arXiv}, we plot the band structure in the original BZ with the weight of the $\bm{q}=\bm{0}$ component in Fig.~\ref{fig5}(d). 
Corresponding to the bare band structure without the spin order (Fig.~\ref{fig4}), the VB top and CB bottom have large weights around the $\Gamma$ and M points, respectively. 
However, the replica band of the VB top located at the M point obtains a weight below the Fermi level due to band hybridization caused by the triple-$\bm{q}$ order of localized spins. 
This weight transfer to the replica band can explain the spectral weight generation around the M point in the ARPES experiment~\cite{ROkuma_arXiv}.

\subsection{RKKY mechanism}

We discuss the origin of the spin order characterized by $\bm{q}_{\rm M}$ [$=\bm{q}_{\alpha}$ in Eq.(\ref{eq:three_q})] based on the RKKY mechanism~\cite{SHayami2021,ZWang2020}. 
Here, we consider the second-order free energy with respect to the Kondo coupling term 
\begin{align}
\hat{H}_{\rm K} \! = \! - \frac{J_{\rm K}}{\sqrt{N}} \sum_{\bm{k},\bm{q}} \sum_{\gamma,m} \sum_{\sigma,\sigma'} \! \left( \hat{d}^{\dag}_{\bm{k}+\bm{q},\gamma, m,\sigma} \bm{\sigma}_{\sigma\sigma'} \hat{d}_{\bm{k},\gamma, m,\sigma'} \right) \! \cdot \! \bm{S}_{\bm{q}, \gamma},
\end{align}
where the Fourier transformation of Eq.~(\ref{eq:H_K}) is conducted with $\hat{c}_{j,\xi} = N^{-\frac{1}{2}} \sum_{\bm{k}} e^{i\bm{k}\cdot \bm{R}_j} \hat{c}_{\bm{k},\xi} $ and $\bm{S}_{j,\gamma} = N^{-\frac{1}{2}} \sum_{\bm{q}} e^{i\bm{q}\cdot \bm{R}_j} \bm{S}_{\bm{q},\gamma}$. 
The second-order free energy of the Kondo lattice model is generally given by $F^{(2)} = -J_K^2 \sum_{\bm{q}} \sum_{\gamma,\gamma'} \sum_{\nu,\nu'} \chi^{\nu \nu'}_{\gamma\gamma'}(\bm{q}) S^{\nu}_{\bm{q},\gamma} S^{\nu'}_{-\bm{q},\gamma'}$, where $\chi^{\nu\nu'}_{\gamma\gamma'}(\bm{q})$ represents the correlation function between the $\nu$ and $\nu'$ components of $d$-electron spins. 
In particular, when the spin part is nearly isotropic [i.e., $\chi^{xx}_{\gamma\gamma'}(\bm{q}) \sim \chi^{yy}_{\gamma\gamma'}(\bm{q}) \sim \chi^{zz}_{\gamma\gamma'}(\bm{q})$], the spin interaction between localized $f$ spins can be characterized by 
\begin{align}
F^{(2)} = -J_{\rm K}^2 \sum_{\bm{q}} \sum_{\gamma,\gamma'} \chi^{\rm (s)}_{\gamma\gamma'}(\bm{q}) \bm{S}_{\bm{q},\gamma} \cdot \bm{S}_{-\bm{q},\gamma'}, 
\end{align}
where $\chi^{\rm (s)}_{\gamma\gamma'}(\bm{q})$ is the spin susceptibility of $d$ electrons. 
This formula suggests that a stable magnetic structure of localized spins can be determined by the configuration (susceptibility) of $d$ electrons near $E_F$. 
While this assessment can suggest a spin order characterized by a modulation wave vector $\bm{q}$, the higher-order free energies should be considered if the stabilization of a triple-$\bm{q}$ order rather than a single-$\bm{q}$ order is justified within the perturbation theory~\cite{SHayami2021,YAkagi2012,SHayami2017}. 

If $d$ electrons are free particles, $\chi^{\rm (s)}_{\gamma\gamma'}(\bm{q})$ is replaced by 
\begin{align}
\chi^{(0)}_{\gamma\gamma'}(\bm{q}) = \sum_{m,m'} \chi^{(0)}_{{\gamma \atop m} \!\! {\gamma \atop m} \!\! {\gamma' \atop m'} \!\! {\gamma' \atop m'}} (\bm{q}), 
\end{align}
where $\sum_{m,m'}$ is the sum for the $d$-orbital components and $\chi^{(0)}_{\zeta_1\zeta_2\zeta_3\zeta_4}(\bm{q})$ with $\zeta=\left( {\Lambda \atop l} \right)$ is the bare susceptibility 
\begin{align}
\chi^{(0)}_{\zeta_1\zeta_2\zeta_3\zeta_4}(\bm{q}) 
&= -\frac{1}{N} \sum_{\bm{k}} \sum_{v,w} 
u_{\zeta_4 v}(\bm{k}+\bm{q}) u^*_{\zeta_2 v}(\bm{k}+\bm{q}) 
\notag \\
\times & u_{\zeta_1 w}(\bm{k}) u^*_{\zeta_3 w}(\bm{k}) 
\frac{f(\varepsilon_v (\bm{k}+\bm{q}))- f(\varepsilon_{w} (\bm{k}))}{\varepsilon_v (\bm{k}+\bm{q})-\varepsilon_{w} (\bm{k})}. 
\label{eq:chi_q_0}
\end{align} 
$\varepsilon_v(\bm{k})$ is the energy of band $v$ in the $d$-$p$ Hamiltonian $\hat{H}^{(0)}_{dp} = \sum_{\bm{k},v,\sigma} \varepsilon_v(\bm{k}) \hat{\varphi}^{\dag}_{\bm{k},v,\sigma} \hat{\varphi}_{\bm{k},v,\sigma}$, which is obtained with diagonalization of Eq.~(\ref{eq:Ham_0_dp}) using the transformation $\hat{c}_{\bm{k},\zeta,\sigma} = \sum_{v} u_{\zeta v}(\bm{k}) \hat{\varphi}_{\bm{k},v,\sigma}$. 
$f(\varepsilon)$ is the Fermi distribution function. 

The formula of $\chi^{(0)}(\bm{q})$ does not take into account correlation effects of $d$ electrons on the spin susceptibility. 
As shown in Fig.~\ref{fig4}(b), both VB top and CB bottom contain the $d$-orbital components, where local Coulomb interactions in Gd $5d$ orbitals, which act as interband interactions, potentially modify the spin susceptibility of $d$ electrons. 
To incorporate correlation effects on the spin susceptibility, we consider the Kanamori interaction term for $d$ electrons $\hat{H}_{U_d}$ (i.e., on-site interactions in the multiorbital Hubbard model) in the random phase approximation (RPA). 
The spin susceptibility $\chi^{({\rm s})}_{\zeta_1\zeta_2\zeta_3\zeta_4}(\bm{q})$ in the RPA is given by 
\begin{align}
\chi^{({\rm s})}(\bm{q}) = \frac{\chi^{(0)}(\bm{q})}{I-U^{({\rm s})}\chi^{(0)}(\bm{q})} , 
\label{eq:RPA_chi_s}
\end{align}
where $U^{({\rm s})}$ represents the vertex of the spin channel 
\begin{align}
U^{({\rm s})}_{m_1m_2m_3m_4} =
\begin{cases}
U_d & (m_1=m_2 = m_3=m_4) \\
U'_d & (m_1=m_3 \ne m_2=m_4) \\
J_d & (m_1=m_2 \ne m_3=m_4) \\
J'_d & (m_1=m_4 \ne m_2=m_3) 
\end{cases}
\end{align}
for the Gd $d$ electrons (where $\gamma_1 = \gamma_2 = \gamma_3 = \gamma_4$). 
$U_d$ and $U'_d$ are the intraorbital and interorbital Coulomb repulsions, respectively, while $J_d$ and $J'_d$ are the strengths of Hund's coupling and pair hopping, respectively. 
Here, we impose $U'_d = U_d - 2J_d$ and $J'_d = J_d$ on the interactions. 

\begin{figure}[t]
\begin{center}
\includegraphics[width=\columnwidth]{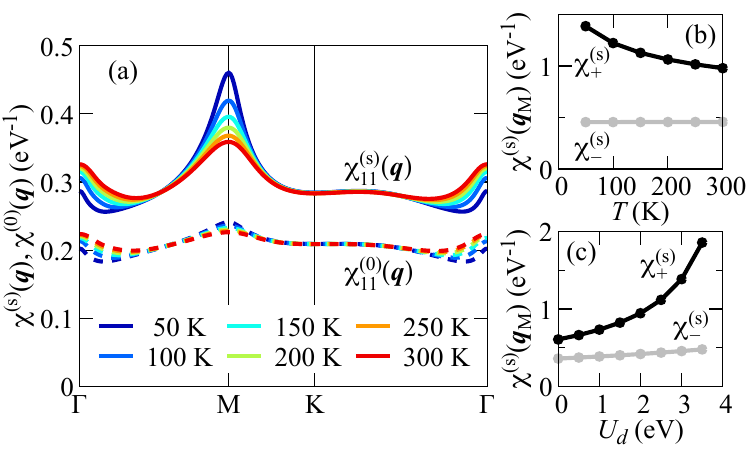} 
\caption{(a) Bare susceptibility $\chi^{(0)}_{11}(\bm{q})$ and spin susceptibility $\chi^{({\rm s})}_{11}(\bm{q})$ in the RPA for $U_d=3$~eV, where $U'_d = U_d - 2J_d$ and $J_d = J'_d=0.2U_d$. 
(b) Temperature dependence of $\chi^{\rm (s)}_{\pm}(\bm{q}_{\rm M})$ at the M point $\bm{q}_{\rm M}$ for $U_d=3$~eV. 
(c) $U_d$ dependence of $\chi^{\rm (s)}_{\pm}(\bm{q}_{\rm M})$ at $T=50$~K. 
$300\times300$ $k$ points are used to calculate $\chi^{(0)}(\bm{q})$.} 
\label{fig6}
\end{center}
\end{figure}

Figure~\ref{fig6} presents the calculated susceptibilities. 
The bare susceptibility $\chi^{(0)}_{\gamma\gamma}(\bm{q})$ has a peak at the M point, whereas its height relative to other $\bm{q}$ points is not so prominent. 
This is because the band gap is nearly zero in our $d$-$p$ model shown in Fig.~\ref{fig4}, and Fermi-surface nesting does not work significantly for the bare susceptibility. 
Hence, the bare susceptibility appears to be weak to develop an order of $f$ spins characterized by $\bm{q}_{\rm M}$. 
Figure~\ref{fig6}(a) also shows the corrected spin susceptibility $\chi^{({\rm s})}_{\gamma\gamma}(\bm{q})$ in the RPA. 
Due to the correlation effects incorporated by the RPA, the peak at the M point becomes prominent at low temperatures. 
Since the value of $\chi^{({\rm s})}(\bm{q}) $ at the M point is large relative to values at other $\bm{q}$ points, the spin susceptibility incorporating the correlation effects between $d$ electrons can support an order of $f$ spins characterized by $\bm{q}_{\rm M}$. 
These results suggest that the Coulomb interactions acting on $d$ electrons near the Fermi level can play a key role in stabilizing a spin order characterized by $\bm{q}_{\rm M}$, such as the all-out spin structure shown in Fig.~\ref{fig5}(a). 

When we assume that spins in the top and bottom sites are parallel ($+$) or antiparallel ($-$), i.e., $\bm{S}_{\bm{q},1} = \pm \bm{S}_{\bm{q},2} = \bm{S}_{\bm{q}}$, the spin structures for $\bm{q}_{\rm M}$ are characterized by $-J_{\rm K}^2 \chi^{\rm (s)}_{\pm}(\bm{q}_{\rm M}) \bm{S}_{\bm{q}_{\rm M}} \cdot \bm{S}_{-\bm{q}_{\rm M}}$ with $\chi^{\rm (s)}_{\pm}(\bm{q}) =\chi^{\rm (s)}_{11}(\bm{q}) +\chi^{\rm (s)}_{22}(\bm{q})\pm [ \chi^{\rm (s)}_{12}(\bm{q}) +\chi^{\rm (s)}_{21}(\bm{q}) ]$. 
Figure~\ref{fig6}(b) shows the temperature dependencies of $\chi^{\rm (s)}_{+}(\bm{q}_{\rm M})$ and $\chi^{\rm (s)}_{-}(\bm{q}_{\rm M})$. 
In our model study, since the spin vertex $U^{({\rm s})}$ is treated as a model parameter, we also plot the $U_d$ dependencies of $\chi^{\rm (s)}_{+}(\bm{q}_{\rm M})$ and $\chi^{\rm (s)}_{-}(\bm{q}_{\rm M})$ in Fig.~\ref{fig6}(c). 
As expected in Eq.~(\ref{eq:RPA_chi_s}), $U_d$ enhances $\chi^{\rm (s)}_{\pm}(\bm{q}_{\rm M})$. 
$\chi^{\rm (s)}_{+}(\bm{q}_{\rm M})$ increases with decreasing temperature and is larger than $\chi^{\rm (s)}_{-}(\bm{q}_{\rm M}) $, suggesting that spins in the top and bottom sites prefer to be parallel at low temperatures. 
This assessment supports the magnetic structure shown in Fig.~\ref{fig5}(a), where spins in the top and bottom layers are parallel.

\section{Summary and outlook} \label{sec_IV}

We investigated the structural, electronic, and magnetic properties of GdGaI. 
We presented the phonon dispersion with no imaginary phonons, suggesting that purely phonon-driven phase transitions are unlikely to occur. 
The electronic band calculation showed that the VBs and CBs near $E_F$ are composed of Gd $5d$ and Ga $4p$ orbitals. 
Then, we constructed a TB model based on the Gd $5d$ and Ga $4p$ orbitals and introduced Kondo coupling between Gd $5d$ electrons and localized spins in Gd $4f$ orbitals. 
The calculated band structure under the magnetic order of localized spins revealed that the flatness of the VB top observed by ARPES in the low-temperature phase can be interpreted by the presence of the triple-$\bm{q}$ spin order. 
Finally, we discussed the origin of the spin order based on the RKKY mechanism. 
We suggested that local Coulomb interactions of Gd $d$ electrons acting as interband interactions near the Fermi level can support the ordering of localized spins. 

Finally, we comment on open issues and future perspectives. 
The DFT bands used for the TB model do not include spin-orbit coupling. 
As presented in Appendix~\ref{appendix_E}, the DFT calculation with spin-orbit coupling barely changes the structures of the VB and the CB near $E_F$, implying that our main scenario may not be significantly modified by spin-orbit coupling. 
If the stability of spin orders is fully evaluated within first-principles calculations, one should perform band calculations that introduce Gd $4f$ orbitals with spin polarization. 
A comprehensive discussion of the energy stability of the ordered phase will be an important issue for GdGaI in the future. 
If we use the Kondo lattice model, evaluating the coupling constant $J_{\rm K}$ is also an important issue. 
Since the size of the band gap depends on the value of $J_{\rm K}$, evaluations of the gap size through experiments, such as scanning tunneling spectroscopy, will provide valuable information. 
In Ref.~\cite{ROkuma_arXiv}, x-ray magnetic circular dichroism (XMCD) was employed to examine the tendency of spin splitting of electronic bands in the magnetic phase. 
While the electronic bands shown in Fig.~\ref{fig5} are doubly degenerate, the presence of a net magnetization of localized spins lifts this degeneracy and gives rise to spin splitting via $J_{\rm K}$. 
Also, spin-orbit coupling brings spin anisotropy, potentially resulting in weak spin splitting of electronic bands under a noncoplanar spin order~\cite{JZhou2016}. 
Careful analysis of XMCD data may also provide meaningful information for $J_{\rm K}$. 
When the all-out spin order shown in Fig.~\ref{fig5}(a) is realized in GdGaI, the spontaneous Hall effect attributed to scalar spin chirality potentially occurs as investigated in the simple triangular Kondo lattice model~\cite{IMartin2008,YAkagi2010,YKato2010}. 

If we discuss GdGaI as a candidate for an excitonic insulator, several issues should be noted. 
In contrast to candidate materials that undergo structural phase transitions, such as TiSe$_2$~\cite{FDiSalvo1976,HCercellier2007} and Ta$_2$NiSe$_5$~\cite{FDiSalvo1986,YWakisaka2009,KSeki2014,YLu2017}, the phonon dispersion of GdGaI shown in Fig.~\ref{fig2}(b) has no imaginary phonons, i.e., structural phase transitions are unlikely to occur in GdGaI. 
This aspect is favorable in discussing the possibility of an excitonic insulator state because electron-phonon coupling cannot be the primary cause of the gap opening. 
On the other hand, Kondo coupling between Gd $5d$ and $4f$ orbitals can drive the gap opening due to the ordering of localized spins (see Fig.~\ref{fig5}) because both the VB top and CB bottom contain the $d$-orbital components [see Fig.~\ref{fig4}(b)]. 
If Kondo coupling alone contributes to the gap opening, there would be no need to consider GdGaI as an excitonic insulator. 
The key issue is the contribution of interband Coulomb interactions to the phase transition. 
In narrow-gap semiconductors (or semimetals with a small band overlap), Fermi-surface nesting, e.g., used in the triangular Kondo lattice model~\cite{IMartin2008}, cannot simply induce a phase transition because the size of the Fermi surface is zero (or small). 
On the other hand, the interband Coulomb interactions can modify the spin susceptibility of $d$ electrons that are mediators of interactions between localized $f$ spins. 
As shown in Fig.~\ref{fig6}, the local $d$-$d$ Coulomb interaction that serves as the interband interactions enhances the spin susceptibility and supports an order of localized spins via the RKKY interaction. 
This suggests that the interband interactions can positively contribute to the phase transition in GdGaI. 
To study the Coulomb contribution to the gap opening in GdGaI, we should perform a self-consistent calculation, such as the Hartree-Fock approximation, incorporating both the Hubbard interaction and Kondo coupling. 
The precise investigation of the Coulomb contribution to the phase transition in GdGaI would be an important issue to discuss this material in the context of the excitonic insulator.

\begin{acknowledgments}
We thank Y.~Okada and R.~Okuma for introducing GdGaI and motivating this work. 
We also thank K.~Aido, D. Gole\v{z}, N.~Jiang, T.~Kondo, K.~Kuroki, A.~J.~Millis, Y.~Murakami, Y.~Ohta, Z.~Sun, S.~Takayoshi, and N.~Tsuji for fruitful discussions. 
This work was supported by Grants-in-Aid for Scientific Research from JSPS, KAKENHI Grant No.~JP20H01849, No.~JP24K06939, No.~JP24H00191, and No.~JP24K01333.
R.M., S.K., and M.O. were supported by JST FOREST Program, Grant No. JPMJFR212P. 
\end{acknowledgments}

\appendix 

\section{Phonon dispersion and electronic band structure using the PBE functional} \label{appendix_A}

\begin{table}[b]
\centering
\caption{Optimized atomic coordinates using the PBE and HSE functionals, shown together with the experimental values.} 
\begin{tabular}{c c c c c c c}
\hline \hline
Atom & \begin{tabular}[c]{@{}c@{}}Wyckoff\\ position\end{tabular} & $x$ & $y$ & $z$ (expt.)\footnote{Experimental values taken from Ref.~\cite{ROkuma_arXiv}} & $z$~(PBE)& $z$ (HSE) \\ 
\hline
Gd & 2c & 0   & 0   & 0.1781(1) & 0.1765 & 0.1753 \\
Ga & 2d & 2/3 & 1/3 & 0.0273(3) & 0.0278 & 0.0293 \\
I  & 2d & 1/3 & 2/3 & 0.3528(2) & 0.3507 & 0.3478 \\ 
\hline \hline
\end{tabular}
\label{table1}
\end{table}

\begin{figure}[t]
\begin{center}
\includegraphics[width=\columnwidth]{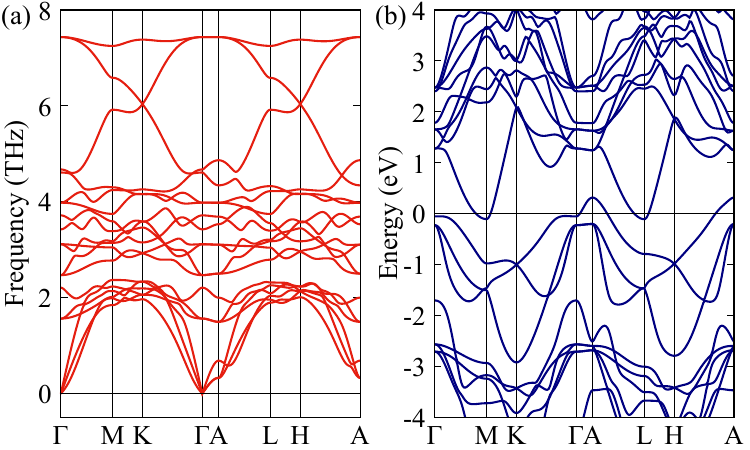} 
\caption{(a) Phonon dispersion and (b) electronic band structure obtained with the PBE functional.} 
\label{fig7}
\end{center}
\end{figure}

We present the results obtained with the PBE functional. 
While plane-wave cutoff energy of 400~eV and an $8 \times 8 \times 3$ $k$ mesh were used in the presented figures, the result exhibits sufficient convergence to that obtained using a larger cutoff energy and a finer $k$ mesh. 
For phonon calculations, we optimized the atomic coordinates. 
Table~\ref{table1} presents the optimized atomic coordinates using the PBE and HSE functionals alongside the experimental values taken from Ref.~\cite{ROkuma_arXiv}. 
Figure~\ref{fig7}(a) shows the phonon dispersion in the optimized crystal structure using PBE. 
Even when PBE is used, the phonon calculation does not exhibit an imaginary phonon. 
The electronic band structure using PBE is presented in Fig.~\ref{fig7}(b), where we used the crystal structure experimentally determined in Ref.~\cite{ROkuma_arXiv}.
Similarly to the result of the HSE functional, the band structure is semimetallic, where the band overlap is larger than that using the HSE functional. 
These results suggest that a band structure with a larger band overlap does not cause phonon softening.

\section{$k_z$ dispersion in bulk GdGaI} \label{appendix_B}

\begin{figure}[t]
\begin{center}
\includegraphics[width=0.94\columnwidth]{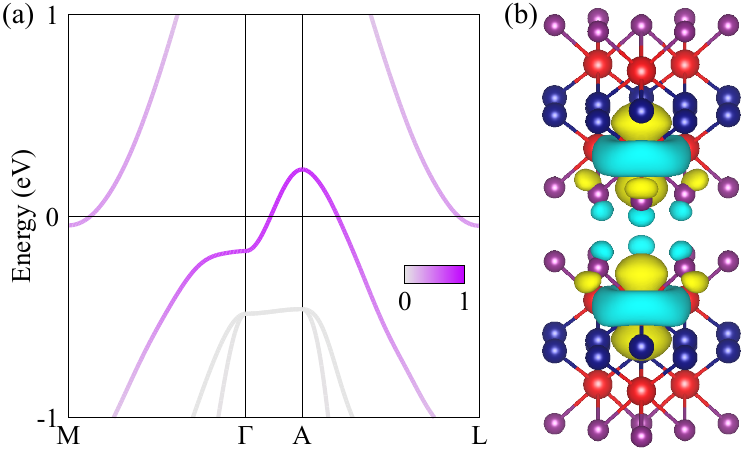} 
\caption{(a) Electronic band structure near $E_F$ of bulk GdGaI. 
(b) Gd $d_{3z^2-r^2}$-based Wannier orbitals that generate interlayer hopping. 
The color bar in (a) indicates the weight of the orbitals shown in (b).} 
\label{fig8}
\end{center}
\end{figure}

We consider the $k_z$ dispersion of the highest VB in bulk GdGaI. 
Figure~\ref{fig8}(a) presents the electronic bands near $E_F$, where the TB model extracted from the DFT bands with the HSE functional is used to plot the band structure. 
As seen in Fig.~\ref{fig8}(a), the $k_z$ dispersion along the $\Gamma$--A line exists near $E_F$ in bulk GdGaI. 
The color on the bands indicates the weight of the Gd $d_{3z^2-r^2}$ like Wannier orbital elongated along the $c$-axis direction [Fig.~\ref{fig8}(b)], revealing that the $d_{3z^2-r^2}$-based orbitals mainly configure the $k_z$ dispersion of the highest VB. 
As shown in Fig.~\ref{fig8}(b), the $d_{3z^2-r^2}$-based Wannier orbitals, including the I orbital components, generate interlayer hopping across the vdW gap. 
The value of the interlayer hopping between these orbitals, obtained by Wannierization, is $\sim0.05$~eV. 
This $k_z$ dispersion disappears as the $c$-axis length of the crystal structure is extended. 
In slab calculations, a sufficiently long $c$-axis length is used to eliminate the $k_z$ dispersion.

\section{Band structures in a single bilayer} \label{appendix_C}

\begin{figure}[t]
\begin{center}
\includegraphics[width=\columnwidth]{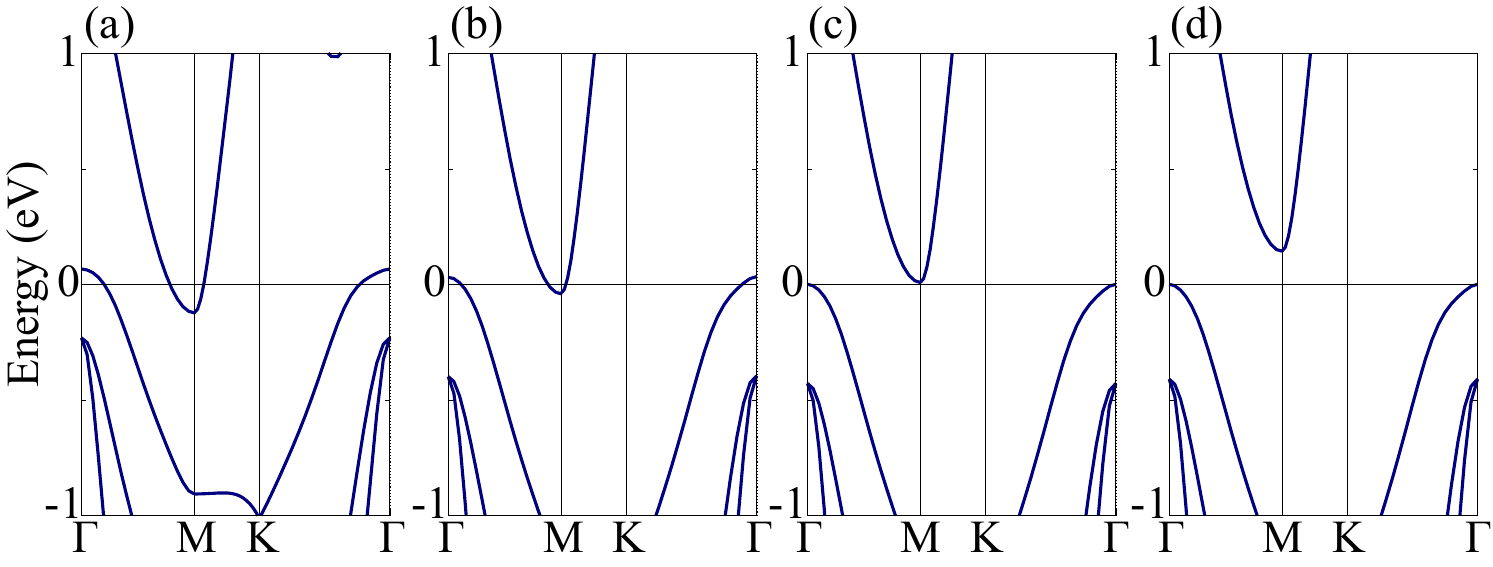} 
\caption{Band structures of single-bilayer GdGaI 
obtained with the (a) PBE and (b)--(d) HSE hybrid functionals. 
The range-separation parameters $\mu=0.2$, 0.15, and 0.1 \AA$^{-1}$ are used in (b), (c), and (d), respectively.} 
\label{fig9}
\end{center}
\end{figure}

\begin{figure}[t]
\begin{center}
\includegraphics[width=\columnwidth]{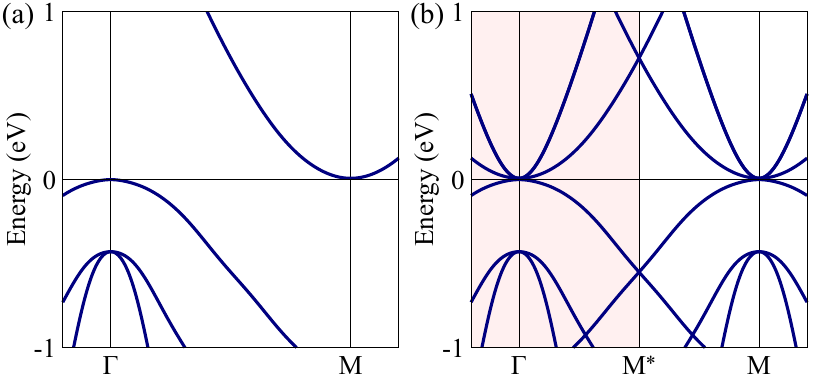} 
\caption{(a) Band structure along the $\Gamma$--M line in the TB model for single-bilayer GdGaI. 
(b) Band folding for the 2$\times$2 superlattice, where Kondo coupling is zero. 
The shadowed area indicates the reduced BZ.} 
\label{fig10}
\end{center}
\end{figure}

We compare the PBE and HSE band structures of a single bilayer. 
Figure~\ref{fig9}(a) shows the band dispersion near $E_F$ obtained by the PBE functional, while Figs.~\ref{fig9}(b)--\ref{fig9}(d) show the band structures obtained by the HSE hybrid functional with different values of the range-separation parameters $\mu$. 
The band structure using PBE is semimetallic. 
When HSE is used, the band overlap is suppressed, and the band gap opens as $\mu$ decreases. 
Since semiconducting behaviors have been suggested experimentally in the normal state~\cite{ROkuma_arXiv}, we employ the nearly zero-gap band structure obtained with $\mu= 0.15$~\AA$^{-1}$ shown in Fig.~\ref{fig9}(c) as the reference DFT band structure for our TB model. 

Figure~\ref{fig10} shows the band structures in the TB model extracted from the DFT bands using HSE with $\mu= 0.15$~\AA$^{-1}$. 
While the bands under the original periodicity appear as shown in Fig.~\ref{fig10}(a), the band folding for the 2$\times$2 period results in the band structure shown in Fig.~\ref{fig10}(b). 
The CB bottoms at the M points are folded to the $\Gamma$ point in the reduced BZ [shadowed area in Fig.~\ref{fig10}(b)]. 
The widely dispersed CB around the $\Gamma$ point is attributed to the CB at the M point ($\bm{k}=\bm{q}_1$) on the indicated $\Gamma$--M line, while the narrowly dispersed CBs are attributed to the CBs at the other two M points ($\bm{k}=\bm{q}_2$ and $\bm{k}=\bm{q}_3$). 
When a 2$\times$2 periodic spin structure, such as the all-out spin structure in Fig.~\ref{fig5}(a), couples to $d$ electrons via Kondo coupling, deformation of the band structure occurs in the reduced BZ, as shown in Fig.~\ref{fig5}(c). 

\begin{figure}[t]
\begin{center}
\includegraphics[width=\columnwidth]{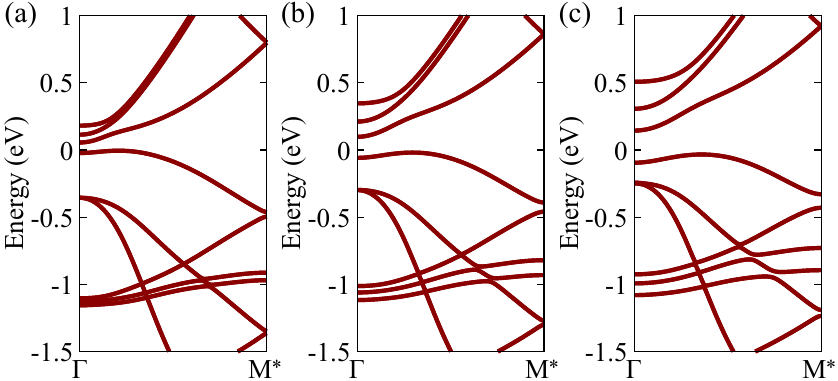} 
\caption{Band structures with the all-out spin structure [Fig.~\ref{fig5}(a)] for (a) $J_{\rm K}=$~0.2~eV, (b) 0.4~eV, and (c) 0.6~eV. 
The zero energy of each figure is set at the lower edge of the band gap.} 
\label{fig11}
\end{center}
\end{figure}

\begin{figure}[t]
\begin{center}
\includegraphics[width=0.95\columnwidth]{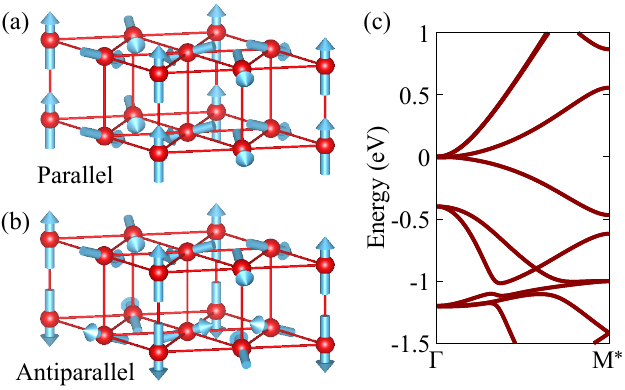} 
\caption{Spin structures with (a) $\eta_1=\eta_2=1$ (parallel) and (b) $\eta_1=-\eta_2=1$ (antiparallel). 
Red balls represent Gd atoms. 
Arrows represent localized spins on Gd $4f$ orbitals. 
(c) Band structure with the antiparallel spin structure in (b), where the electronic bands for $J_{\rm K}=0.4$~eV are plotted in the reduced BZ.} 
\label{fig12}
\end{center}
\end{figure}

\begin{figure}[t]
\begin{center}
\includegraphics[width=0.95\columnwidth]{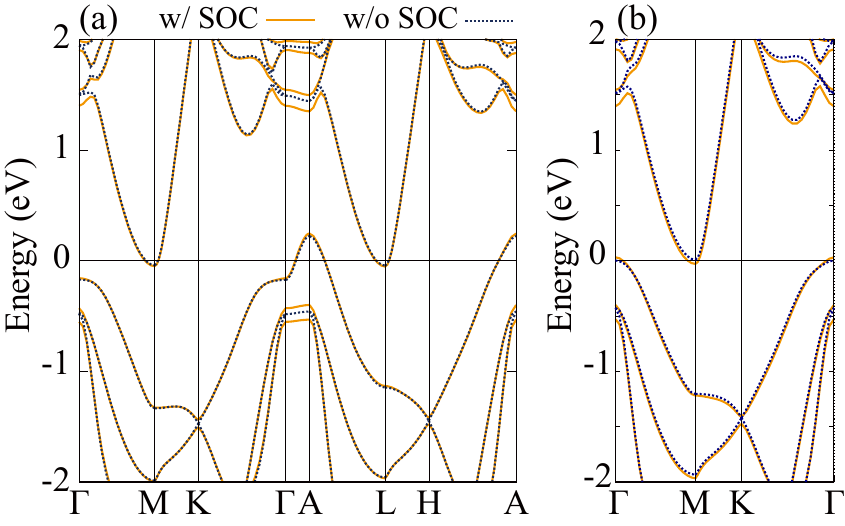} 
\caption{Band structures of (a) bulk GdGaI and (b) single-bilayer GdGaI with spin-orbit coupling. 
The solid and dotted lines are the band dispersions with and without spin-orbit coupling, respectively.} 
\label{fig13}
\end{center}
\end{figure}

\section{All-out spin structure} \label{appendix_D}

In Sec.~\ref{sec_IIIB}, we show the band structure that assumes all-out order of localized spins for a single $J_{\rm K}$ value. 
Here, we present the band dispersions for various $J_{\rm K}$ values. 
The bands are folded into the reduced BZ as shown in Fig.~\ref{fig10}(b), and the VB and CB form a hybridization gap at the $\Gamma$ point via Kondo coupling $J_{\rm K}$. 
As shown in Fig.~\ref{fig11}, the gap size increases monotonically with $J_{\rm K}$, indicating that Kondo coupling is a key factor of gap opening in the all-out spin structure. 

In the above discussions, we consider the spin structure shown in Fig.~\ref{fig12}(a), where localized spins in the top and bottom layers are parallel ($\eta_1=\eta_2=1$). 
On the other hand, antiparallel spins in the top and bottom layers ($\eta_1=-\eta_2=1$) form the spin structure shown in Fig.~\ref{fig12}(b). 
Figure~\ref{fig12}(c) shows the electronic bands for the spin structure in Fig.~\ref{fig12}(b). 
In contrast to the band structure for the all-out spin structure shown in Fig.~\ref{fig11}, the band gap near $E_F$ is not opened when spins are antiparallel. 
Due to this difference, the energy of the spin structure in Fig.~\ref{fig12}(a) is lower than that in Fig.~\ref{fig12}(b).

\section{Effects of spin-orbit coupling} \label{appendix_E}

Figure~\ref{fig13} presents the band structures with and without spin-orbit coupling, where the band dispersions obtained with the HSE hybrid functional ($\mu=0.15$~\AA$^{-1}$) are plotted. 
The band dispersions of bulk GdGaI and single-bilayer GdGaI are plotted in Figs.~\ref{fig13}(a) and ~\ref{fig13}(b), respectively. 
While there are changes in the band structures away from the Fermi level $E_F$, the VB and CB near $E_F$ are not substantially modified by spin-orbit coupling. 
Therefore, the band structure near $E_F$, which can strongly contribute to the phase transition, remains even when spin-orbit coupling is introduced into the band calculation.

\bibliography{reference}

\end{document}